\documentclass[preprint,prb,showpacs]{revtex4-1}
\usepackage{graphicx}
\usepackage{bm}
\usepackage{natbib}
\usepackage{color}

\begin{document}
\preprint{APS/123-QED}
\title{Oxygen vacancy-induced structural evolution of SrFeO$_{3-x}$ epitaxial thin film from brownmillerite to perovskite}

\author{Seulki Roh}
\author{Seokbae Lee}
\author{Myounghoon Lee}
\author{Yu-Seong Seo}
\author{Amit Khare}
\author{Taesup Yoo}
\author{Sungmin Woo}
\author{Woo Seok Choi}
\author{Jungseek Hwang}
\email{Electronic address: jungseek@skku.edu}
\affiliation{Department of Physics, Sungkyunkwan University, Suwon, Gyeonggi-do 16419, Republic of Korea}

\author{A. Glamazda}
\author{K.-Y. Choi}
\affiliation{Department of physics, Chung-Ang University, Seoul 06980, Republic of Korea}

\date{\today}

\begin{abstract}
We investigated SrFeO$_{3-x}$ thin films on a SrTiO$_3$ (001) substrate prepared via pulsed laser epitaxy using an optical spectroscopy technique. The oxygen vacancy level ($x$) was controlled by post-annealing processes at different oxygen partial pressures. We achieved a brownmillerite(BM) structure at $x =$ 0.5 and observed the evolution of the crystal structure from BM into perovskite(PV) as the oxygen concentration increased. We observed the evolution of infrared-active phonons with respect to the oxygen concentration, which was closely related to the structural evolution observed via X-ray diffraction. We identified the phonons using the shell-model calculation. Furthermore, we studied temperature-dependent behaviors of the phonon modes of three representative samples: PV, and two BMs (BM$_{\mathrm{oop}}$ and BM$_{\mathrm{ip}}$) with different orientations of the oxygen vacancy channel. In the BM$_{\mathrm{oop}}$ sample, we observed a phonon mode, which exhibited an unusual red-shift with decreasing temperature; this behavior may have been due to the apical oxygen instability in the FeO$_6$ octahedron. Our results provide important information regarding the ionic conduction mechanism in SrFeO$_{3-x}$ material systems.
\end{abstract}

\maketitle

\section{Introduction}

The multivalent nature of the transition-metal oxides gives rise to different oxygen stoichiometries that result in various physical properties \cite{Choi2013,Jeen2013o,Jeen2013n,Khare2017,Khare2017a}. Brownmillerite (BM) is a particular oxygen-deficient perovskite(PV) oxide structure, that consists of alternating stacked layers of tetrahedral and octahedral units, with oxygen vacancy channels (OVC) aligned in the tetrahedral layer. The OVC plays a key role in ionic conduction and/or oxygen storage \cite{Jeen2013o,Piovano2015,Inoue2010,Auckett2013,Young2015}. It has been reported that BM transition-metal oxides, including SrCoO$_{3-x}$ (SCO), exhibit a topotactic phase transition, manifesting their ability to absorb oxygen \cite{Jeen2013t,Jeen2013n}. SCO can be stabilized in two different crystalline phases depending on the oxygen content $x$: BM with $x$ = 0.5 and PV with $x$ = 0. These two phases can be reversibly switched by gaining or losing oxygen. On the other hand, the ionic conduction in BM is highly anisotropic, showing high conductivity along the OVC direction \cite{Inoue2010}. Notably, a different crystal plane of the substrate is adopted to achieve a different orientation of the OVC of the BM films on the substrate because the orthorhombic BM structure has different lattice constants with respect to the crystal axis. By switching the orientation of the OVC in the BM structure, a higher ionic conductivity can be achieved along the OVC, which is tilted 45$^{\circ}$ with respect to the chain of the tetrahedral units \cite{Jeen2013o}. Owing to these interesting unique properties of the BM structure, it can be applied to various electronic devices \cite{Auckett2013,Acharya2016}.

SrFeO$_{3-x}$ (SFO) also shows topotactic phase transitions between BM and PV phases, which can be achieved at relatively low temperatures because their Gibbs free-energy difference is small (on the order of 100 meV) \cite{Khare2017}. SrFeO$_{3}$ has a PV crystal structure and exhibits metallic behavior. SrFeO$_{2.5}$ has a BM crystal structure, comprising alternately stacked FeO$_6$ octahedral and FeO$_4$ tetrahedral layers. The stacking axis is the orthorhombic $b$-axis and the OVCs are along the orthorhombic [101]-direction within the tetrahedral layers\cite{Khare2017a,Khare2017,Haavik2003,Hodges2000} (refer to the middle and bottom figures of Fig. 1(b)). Interestingly, when SrFeO$_{2.5}$ is deposited on a SrTiO$_3$ (STO) (001) substrate via pulsed laser epitaxy (PLE), one can prepare two different SrFeO$_{2.5}$ BM thin film samples depending on the orientation of the oxygen-deficient (FeO$_4$) tetrahedral layer (ODTL) with respect to the film surface; one BM has ODTLs parallel to the film surface or in-plane ODTLs (BM$_{\mathrm{ip}}$) (refer to the middle figure of Fig. 1(b)) and the other BM has ODTLs perpendicular to the film or out-of-plane ODTLs (BM$_{\mathrm{oop}}$) (refer to the bottom figure of Fig. 1(b)). Because the BM$_{\mathrm{ip}}$ and BM$_{\mathrm{oop}}$ films on STO (001) substrates exhibit similar free energies \cite{Young2015}, the two different BM phases of the films are sensitive to their surrounding conditions; the two phases can compete with each other as kinds of order parameters \cite{Young2015,Shimakawa2010}. The existence of these two possible BM phases on a substrate surface provides an opportunity to study not only the phonons in the $ab$-plane but also the phonons along the $c$-axis of the orthorhombic BM structure.

In this study, we prepared a series of SrFeO$_{3-x}$ (SFO) epitaxial thin films on an STO (001) substrate using PLE and subsequent post-annealing processes. The crystal structure of the prepared SrFeO$_{3-x}$ thin film varies from BM ($x =$ 0.5) to PV ($x =$ 0) depending on the oxygen deficiency level ($x$). The oxygen deficiency level can be controlled according to the oxygen partial pressure (OPP) in the post-annealing processes. We performed a X-ray diffraction(XRD) experiment to investigate the evolution of the crystal structure. We used an optical spectroscopy technique to study the infrared (IR)-active phonon modes of the SrFeO$_{3-x}$ thin film samples. We observed the gradual evolution of the phonon modes as a function of the oxygen deficiency level. We also observed that the BM$_{\mathrm{oop}}$ sample exhibited quite different phonon modes from the BM$_{\mathrm{ip}}$ sample, as we measured different facets of the BM crystal. To identify and assign the optical phonon modes, we studied the $\Gamma$-point phonon modes by performing shell-model lattice dynamic calculations. We also cooled three (BM$_{\mathrm{oop}}$, BM$_{\mathrm{ip}}$, and PV) representative samples to examine the temperature-dependent properties of the observed phonon modes. We observed one interesting phonon mode in the BM$_{\mathrm{oop}}$ sample, which exhibited an unusual red-shift with the temperature reduction. The unusual temperature-dependent behavior of this phonon mode is attributed to the instability of the apical oxygen in the octahedron \cite{Paulus2008}.

\section{Sample preparation}

We prepared high-quality epitaxial SrFeO$_{3-x}$ thin film samples ($\approx$ 18 nm thick) on a TiO$_2$-terminated STO substrate (001) using PLE at 700 $^{\circ}$C. To obtain SrFeO$_{3-x}$ thin films, we treated the as-grown SrFeO$_{2.5}$ thin films in the BM$_{\mathrm{oop}}$ phase using a post-annealing process at 700 $^{\circ}$C under different OPPs (0.01 - 500 Torr) for 10 minutes. Our as-grown SrFeO$_{2.5}$ sample shows the BM$_{\mathrm{oop}}$ phase, whose oxygen-deficient tetrahedral layers (ODTLs) are perpendicular to the film surface. However, 10 min of post-annealing under an OPP of 0.01 Torr at 700 $^{\circ}$C changes the orientation to the BM$_{\mathrm{ip}}$ phase, whose ODTLs are parallel to the substrate surface. By adding oxygen to the prepared BM$_{\mathrm{ip}}$ SrFeO$_{2.5}$ thin film using a post-annealing process with various OPPs, we prepared a series of SrFeO$_{3-x}$ thin films, including the PV film. We denote our thin film samples according to their OPPs, which are 0.01, 0.05, 0.1, 0.5, 1, 80, and 500 Torr. Note that we have two different BM structures at the same OPP of 0.01: the as-grown structure (BM$_{\mathrm{oop}}$) and the structure that was post-annealed (BM$_{\mathrm{ip}}$) for 10 min. These two BM films have exactly the same crystal structure; the only difference is their orientations, which differ by 90$^{\circ}$, as shown in Fig. 1(b).

\begin{figure}[ht]
  \vspace*{-0.0 cm}%
  \centerline{\includegraphics[width=4.5 in]{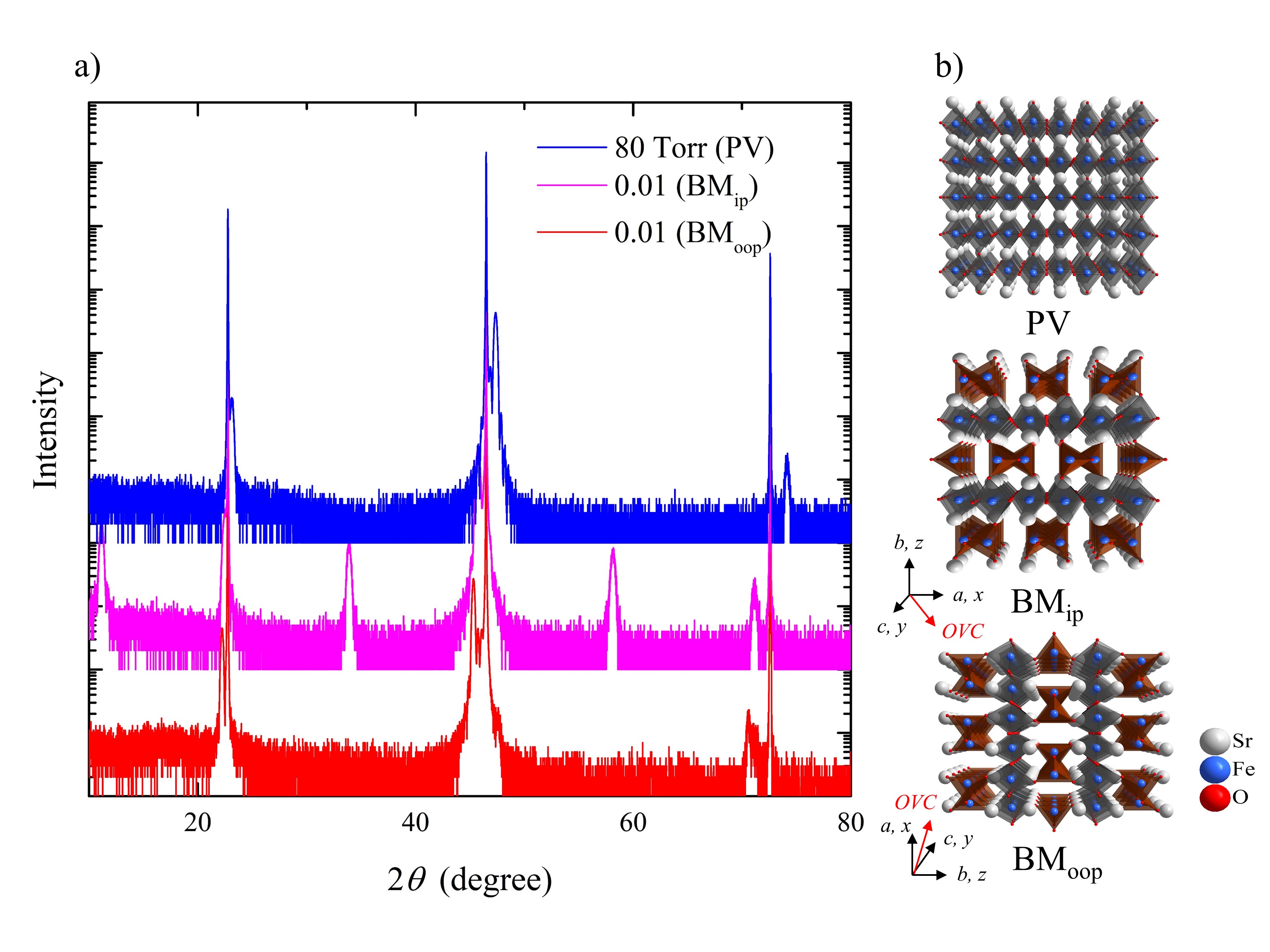}}%
  \vspace*{-0.0 cm}%
\caption{(a) XRD data of three representative SFO films on STO (SFOs/STO) samples (part of the data is reproduced from Ref. \cite{Khare2017}). (b) Three-dimensional crystal structures of the three crystal structures: PV and BM (BM$_{\mathrm{ip}}$ and BM$_{\mathrm{oop}}$). $x$, $y$ and $z$-ases also are marked for discussions in section IV.}
 \label{fig1}
\end{figure}

Fig. 1(a) shows XRD patterns of three representative SFO thin film samples. The lower two patterns show BM$_{\mathrm{oop}}$ (the lowest) and BM$_{\mathrm{ip}}$ (the middle). We can see the half-order peaks between the major peaks in the XRD pattern of the BM$_{\mathrm{ip}}$ sample. The existence of the half-order peaks indicates that the lattice constant parallel to the film surface is doubled by the 90$^{\circ}$ rotation of the orientation. As we increase the OPP further, the lattice constant ceases doubling along the film, and the crystal structure of the thin film transformed into PV-like one, losing the half-order peaks in the XRD pattern. Fig. 1(b) shows side-views of three major crystal structures of our thin films: BM$_{\mathrm{oop}}$, BM$_{\mathrm{ip}}$, and PV. The horizontal substrate surface is located at the bottom of each crystal structure. Additional information regarding the sample preparation can be found in a previous work\cite{Khare2017}.

\section{Optical measurement and data analysis}

Optical spectroscopy is an elegant experimental technique that can reveal the electronic and phononic structures of material systems. Here, we are interested in phonon modes, which are collective excitations of the lattice vibrations and can carry information regarding the crystal structure. Optical phonons usually have characteristic energy scales of far-IR (FIR) and mid-IR (MIR). We obtained reflectance spectra of our SFO thin films on STO substrates in the FIR region (50 - 700 cm$^{-1}$) using an FTIR-type spectrometer (Vertex 80v, Bruker) equipped with a 4.2 K bolometer detector. An $in-situ$ Au-evaporation technique was applied to obtain accurate reflectance spectra \cite{Homes1993}. For the temperature-dependent experiments, the samples were cooled below room temperature using a continuous (liquid N$_2$) flow cryostat. Fig. 2 shows the measured reflectance spectra of all our SFO thin films on STO (SFO/STO samples) along with the bare STO substrate in the FIR region. Because our films are thin ($\sim$18 nm) and transparent in the FIR region, except for the frequencies of characteristic phonon modes, the reflectance spectra of all the SFO/STO samples appear similar to the typical reflectance spectrum of STO. In the FIR range, STO has four IR-active phonon modes: 93 (Slater mode), 176 (Last mode), 436 (appears below 105 K with the tetragonal phase transition), and 548 cm$^{-1}$ (Axe mode) \cite{Petzelt2001,Sirenko2000,Last1957,Fleury1976,Fedorov1998}. We clearly observe the three phonon modes of the STO substrate in the measured reflectance spectra of all the SFO/STO samples; they appear as strong dips near 50, 175, and 480 cm$^{-1}$ in the reflectance. We can also observe small dips at various frequencies, which are the phonon modes of the SrFeO$_{3-x}$ thin films.

\begin{figure}[t]
  \vspace*{-0.3 cm}%
  \centerline{\includegraphics[width=4.5 in]{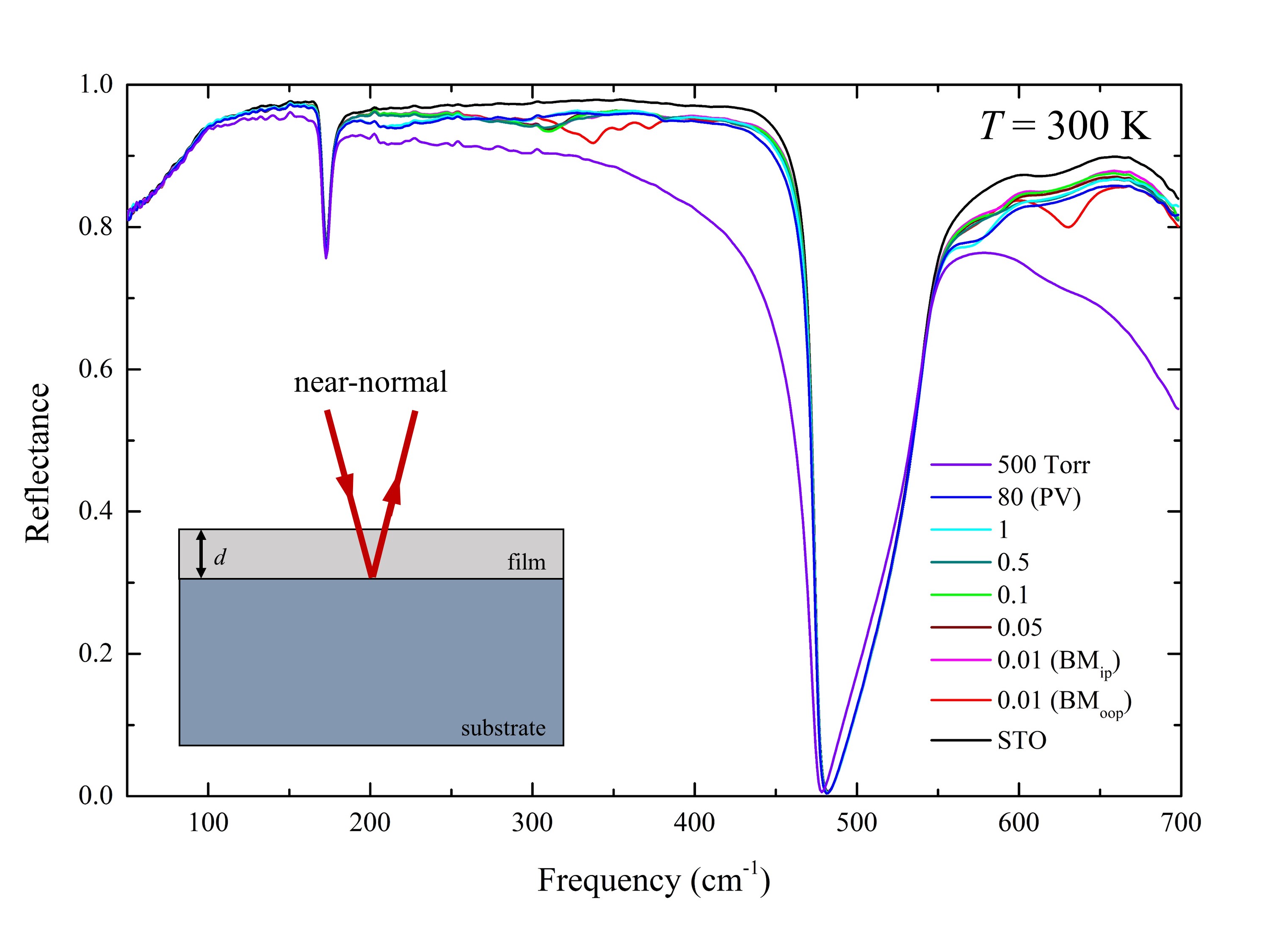}}%
  \vspace*{-0.3cm}%
\caption{Measured reflectance spectra of our all the SFO/STO samples including the sample treated at OPP = 500 Torr and the bare STO substrate. Three strong phonon modes of the STO substrate are dominantly observable. The sample treated at OPP = 500 Torr exhibits a strong Drude absorption over the measured FIR range, while other samples show small dips caused by the phonon absorptions of the films.}
 \label{fig2}
\end{figure}

To separate the phonon modes of the SFO thin films from the measured reflectance spectra of the SFO/STO samples, we exploit the high and flat reflectance of the STO substrate in the FIR range. The available regions are 100 - 430 cm$^{-1}$ ($R(\omega) \approx$ 97\%) and 550 - 700 cm$^{-1}$ ($R(\omega) \approx$ 88\%). By dividing the reflectance of the SFO/STO sample by that of STO in the available regions, we can obtain an {\it effective} transmittance spectrum of the SFO thin film. In this situation, the light from the source effectively passes through the SFO film twice before it arrives at the detector; it first goes though the SFO film, is then bounced back by the high-reflecting STO surface at the interface between the SFO and STO, and then passes through the SFO film a second time, as depicted in the inset of Fig. 2. We emphasize that this approach is reliable whenever the reflectance spectrum of the substrate is high and flat without any sharp features. Note that there are no absorptions by the insulating SrFeO$_{3-x}$ film in the FIR range, except for the absorptions by its phonons. However, as $x$ decreases, the SrFeO$_{3-x}$ film becomes increasingly metallic. A metallic SrFeO$_{3-x}$ film exhibit an additional Drude absorption, as indicated by the reflectance spectrum of the SFO/STO sample treated at OPP = 500 Torr in Fig. 2. The strong Drude absorption governs the whole FIR spectrum and screens the phonon modes almost completely. Therefore, we exclude this SFO sample treated at 500 Torr from our subsequent discussion. We also tried to extract phonon modes using a multilayer model and found that our method described below is better to observe small intensity phonons (refer to Supplementary Material (SM)\cite{sm:2017}).

Having determined the {\it effective} transmittance spectra of the SFO thin films, we can go one step further to obtain the absorption coefficient spectra ($\alpha(\omega)$) of the SFO films using the well-known Beer's law, which can be written as follows:
\begin{equation}\label{eq1}
\alpha(\omega) = -\frac{\ln T_{\mathrm{eff}}(\omega)}{d_{\mathrm{eff}}},
\end{equation}
where $T_{\mathrm{eff}}(\omega)$ is the effective transmittance, and $d_{\mathrm{eff}}$ is the effective thickness of the SFO thin film, which is twice its physical thickness ($d$). After we obtain the absorption coefficient spectra of the SFO films, we fit the absorption coefficient spectra with a simple Lorentz model to find and identify the phonon modes. Here, we note that a Lorentz component corresponds to a phonon mode. The absorption coefficient can be expressed in terms of other optical constants, as follows:
\begin{equation}\label{eq2}
\alpha(\omega) =\frac{ 4 \pi {\sigma_{1}(\omega)}}{n(\omega)},
\end{equation}
where $\sigma_{1}(\omega)$ is the real part of the optical conductivity, and $n(\omega)$ is the index of refraction. The optical conductivity, which is given by $\tilde{\sigma}(\omega) \equiv \sigma_1(\omega)+i\sigma_2(\omega)$, can be written in the Lorentz model as follows:
\begin{equation}\label{eq3}
\tilde{\sigma}(\omega) = \frac{i}{4 \pi} \sum_{j}\frac{\Omega_{p,j}^2 \:\omega}{\omega_{0j}^2-\omega^2+i\gamma_j\omega},
\end{equation}
where $\Omega_{p,j}$, $\omega_{0j}$, and $\gamma_j$ are the plasma frequency, the center frequency, and the width of the $j^{th}$ Lorentz (or phonon) component, respectively. The optical conductivity ($\tilde{\sigma}(\omega)$) is related to the complex index of refraction ($\tilde{N}(\omega) \equiv n(\omega)+i\kappa(\omega)$) as follows: $\tilde{N}(\omega) = \sqrt{i4 \pi \tilde{\sigma}(\omega)/\omega + \epsilon_{\mathrm{H}}}$, where $\kappa(\omega)$ is the extinction coefficient, and $\epsilon_{\mathrm{H}}$ is the high-energy background dielectric constant. Because of the $n(\omega)$ in the denominator of Eq. (2), one additional oscillator centered at $\omega_0$ may affect the absorption spectrum near $\omega_0$; while it lowers the resulting $\alpha(\omega)$ below $\omega_0$, it increases the resulting $\alpha(\omega)$ above $\omega_0$. Therefore, individual modes may exceed the whole fit.

We calculated the $\Gamma$- point phonon modes to assign the symmetries to the observed IR-active phonon modes by performing shell-model lattice dynamical calculations using the General Utility Lattice Program (\textit{GULP}) package \cite{Gale1997}. The symmetry of the calculated modes was analyzed by the Bilbao Crystallographic Server \cite{Aroyo2006}. In the shell-model, the charge of the ion ($Z$) is treated as a combination of the sum of the point core with charge $X$ and the massless shell with charge $Y$ that models the valence electrons. The ionic polarizability $\alpha=Y^2/K$ can be described as a harmonic oscillator with a force constant $K$, which originates from the interaction between the core and the shell. The inter-ionic interactions can be represented as follows, including the long-range Coulomb potentials and short-range Born-Mayer-Buckingham potentials between ions $i$ and $j$:
\begin{equation}\label{eq4}
V_{inter-ion}(r) =A_{ij}\:e^{-(r/\rho_{ij})}-\frac{C_{ij}}{r^6}
\end{equation}
where $A_{ij}$ and $\rho_{ij}$ are the strength and range of the repulsive interaction, respectively, and $C_{ij}$ is an attractive part with the inter-ionic distance $r$. The shell-model parameters are optimized by starting from well-documented data to achieve reasonable agreement with the experimental data \cite{Haavik2003,Damljanovi2008}.

\section{Results and discussion}

The resulting shell-model parameters are summarized in Table I. For comparison, we also performed the shell-model lattice dynamical calculations for SrFeO$_3$. The adopted shell-model parameters are listed in Table II. The three T$_{1u}$ modes are calculated at 172, 249, and 559 cm$^{-1}$ (refer to the top panel of Fig. 3 and Fig. 4). The calculated modes agree reasonably well with the observed modes at 214, 294, and 578 cm$^{-1}$. The calculated eigenvectors of the three IR-active phonon modes are depicted in Fig. 4. To determine the crystal group of BM$_{\mathrm{oop}}$ and BM$_{\mathrm{ip}}$, we performed dynamical calculations for the $Ibm2$, $Pbma$, and $Icmm$ crystal groups. We found that the $Pbma$ space group provides a better description of the experimental data with respect to the phonon energies and orientation dependence. Fig. 3 plots the calculated IR-active phonon modes of the BM$_{\mathrm{oop}}$, BM$_{\mathrm{ip}}$, and PV crystal structures. Here, the BM$_{\mathrm{oop}}$ (BM$_{\mathrm{ip}}$) probes the IR-active modes polarized in the $yz$ ($xy$)-plane. It is worthwhile to note that the $x$, $y$, and $z$ axes correspond to the $a$, $c$, and $b$ orthorhombic crystal axes, respectively (refer to the Fig. 1(b)). The displacement patterns of the representative normal modes are also plotted in Fig. 5.

\begin{figure}[t]
  \vspace*{-0.3 cm}%
  \centerline{\includegraphics[width=4.5 in]{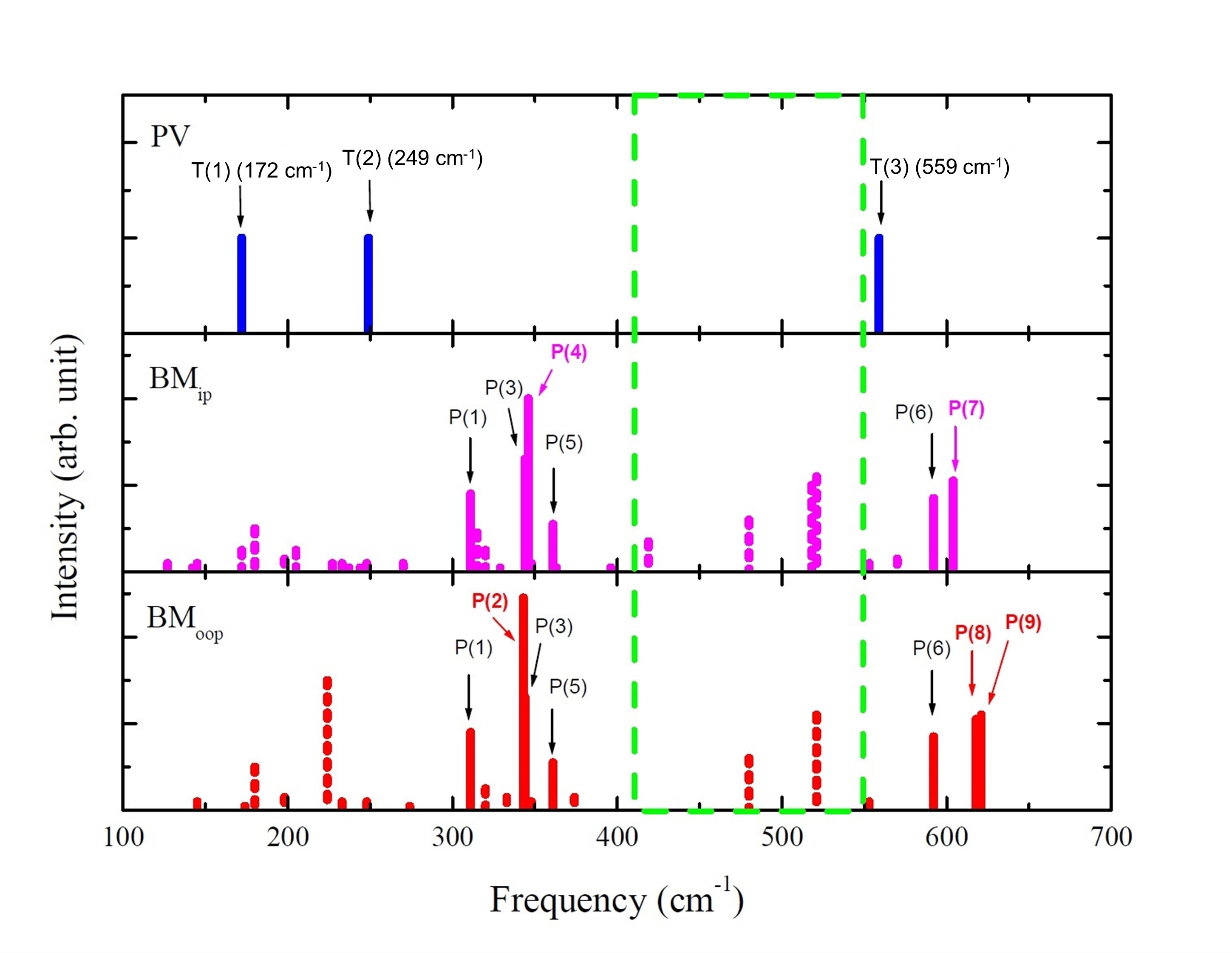}}%
  \vspace*{-0.3cm}%
\caption{Calculated IR-active modes of SrFeO$_3$ ($Pm\bar{3}m$) and SrFeO$_{2.5}$ ($Pbma$) for the BM$_{\mathrm{ip}}$ ($E \Arrowvert x,y$) and BM$_{\mathrm{oop}}$ ($E \Arrowvert y,z$) configurations. The intensity of the phonon peak is calculated using $GULP$, as described in the text. The phonons in the green dashed rectangle are not accessible with our experiment due to the strong phonons of the STO substrate.}
 \label{fig3}
\end{figure}

\begin{figure}[t]
  \vspace*{-0.5 cm}%
  \centerline{\includegraphics[width=4.5 in]{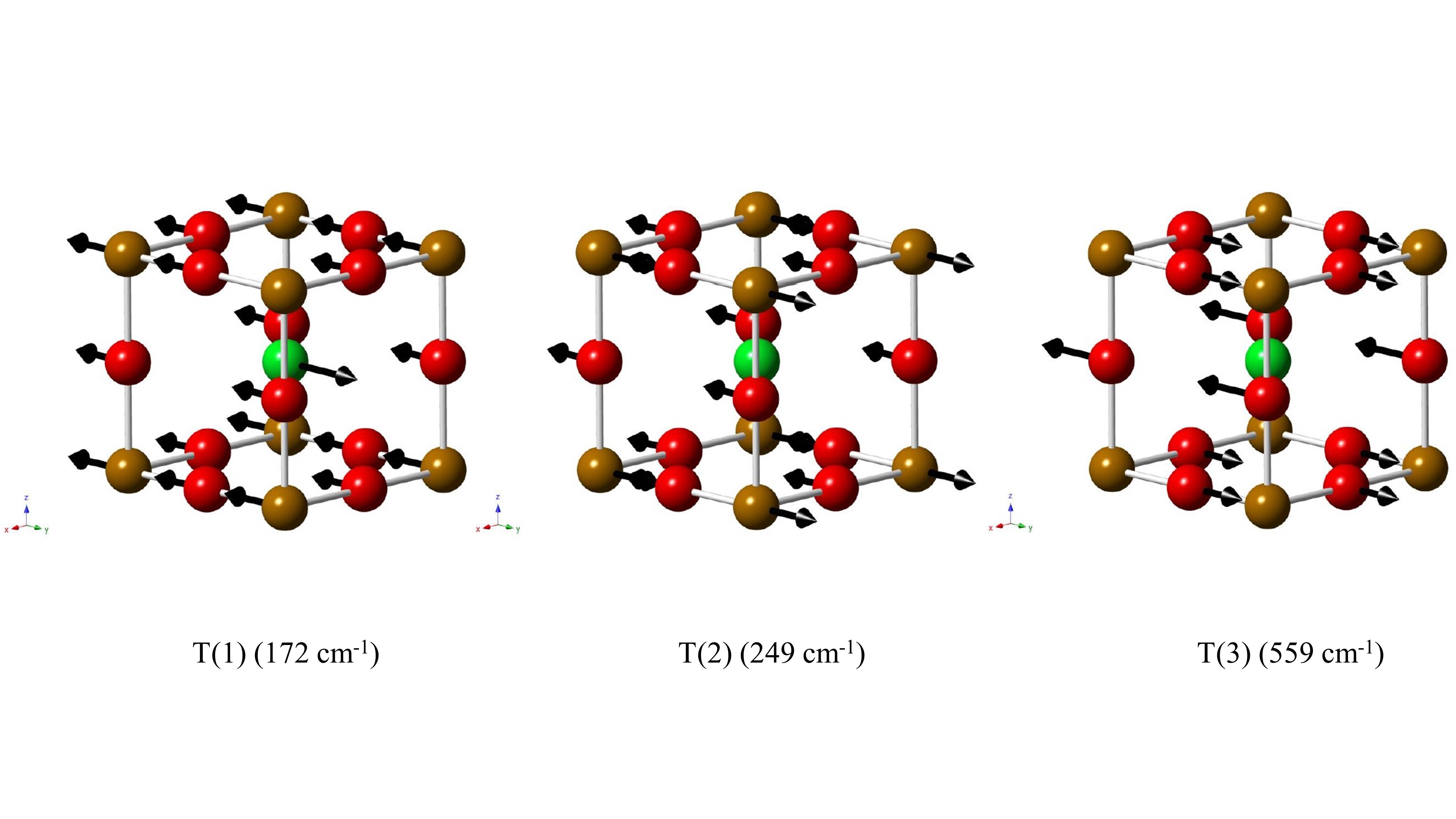}}%
  \vspace*{-1.3cm}%
\caption{Schematic representation of the calculated IR-active normal modes in SrFeO$_3$. The calculated frequencies are shown in parentheses. The amplitude of the vibrations is represented by the length of the arrows. The mustard-colored balls represent Fe, and the red and green balls represent oxygen and strontium atoms, respectively.}
 \label{fig4}
\end{figure}

\begin{table}[]
\centering
\caption{Adopted inter-atomic potential parameter for SrFeO$_{2.5}$.}
\resizebox{\columnwidth}{!}{
\begin{tabular}{|c|c|c|c|c|c|c|}
\hline
Ions & $Y$ (e) & $K$ (eV/$\AA^2$) & Atomic pair & $A_{ij}$ (eV) & $\rho_{ij}$ ($\AA$) & $C_{ij}$ (eV$\AA$) \\ \hline
Fe$^{3+}$ & 1.0    & 990  & Fe$^{3+}$-O$^{2-}$ & 3358.40    & 0.265  & 0  \\ \hline
Sr$^{2+}$ & 1.0    & 30   & Sr$^{2+}$-O$^{2-}$ & 3219.96    & 0.307  & 0  \\ \hline
O$^{2-}$  & -2.5   & 19   & O$^{2-}$-O$^{2-}$  & 249.38     & 0.352  & 0  \\ \hline
\end{tabular}
}
\end{table}

\begin{table}[]
\centering
\caption{Resulting inter-atomic potential parameter for SrFeO$_{3}$.}
\resizebox{\columnwidth}{!}{
\begin{tabular}{|c|c|c|c|c|c|c|}
\hline
Ions & $Y$ (e) & $K$ (eV/$\AA^2$) & Atomic pair & $A_{ij}$ (eV) & $\rho_{ij}$ ($\AA$) & $C_{ij}$ (eV$\AA$) \\ \hline
Fe$^{3+}$ & 1.029  & 915.4 & Fe$^{4+}$-O$^{2-}$ & 1560.0 & 0.299  & 0  \\ \hline
Sr$^{2+}$ & 1.831  & 9.5   & Sr$^{2+}$-O$^{2-}$ & 1435.7  & 0.337 & 0  \\ \hline
O$^{2-}$  & -2.513 & 34.5  & O$^{2-}$-O$^{2-}$  & 22764.0 & 0.149 & 0  \\ \hline
\end{tabular}
}
\end{table}

\begin{figure}[t]
  \vspace*{-1.8 cm}%
  \centerline{\includegraphics[width=5.0 in]{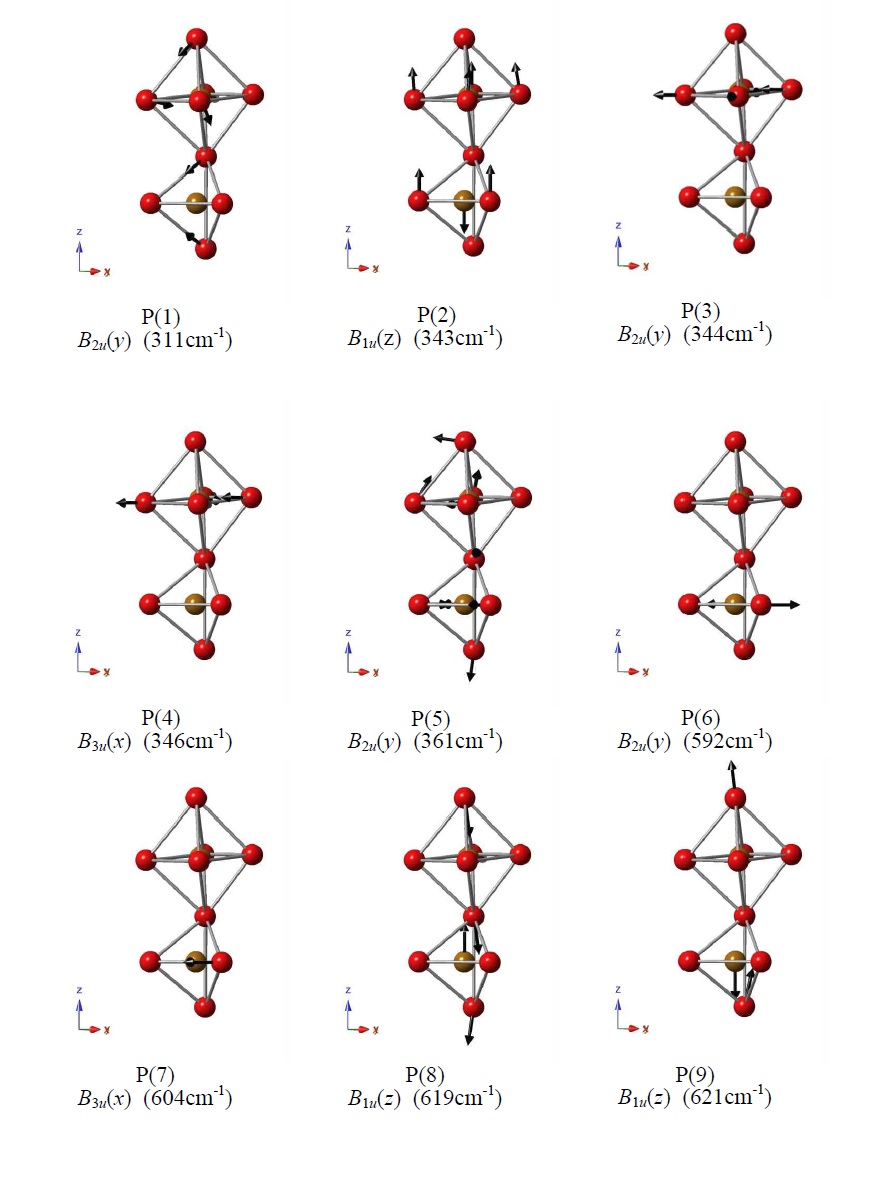}}%
  \vspace*{-1cm}%
\caption{Schematic representation of the most intensive IR-active modes P1-P9 with calculated frequencies in the brackets. The related amplitude of the vibrations is represented by the arrow length. The mustard-colored balls represent Fe(1)/Fe(2) and Fe(3)/Fe(4) ions, which belong to the octahedron and tetrahedron, respectively. The red balls indicate the equatorial octahedron O(1)/O(2), tetrahedral O(5)/O(6,) and apical O(3)/O(4). The modes including strontium vibrations are observed in the low frequency region and have weak intensities. The strontium atoms are not shown, for clarity.}
 \label{fig5}
\end{figure}

\begin{table}[]
\centering
\caption{List of phonon symmetries and, experimental and calculated frequencies at the $\Gamma$-point for SrFeO$_{2.5}$ with $Pbma$ symmetry.}
\resizebox{\columnwidth}{!}{
\begin{tabular}{|c|c|c|c|}
\hline
mode & Expt(cm$^{-1}$) & Calc (cm$^{-1}$) & Assignment                                         \\ \hline
P(1) & 282 & 311  & $B_{2u}$($y$) (FeO$_4$ and FeO$_6$ bending vibrations)                     \\ \hline
P(2) & {\bf 329}  & 343 & $B_{1u}$($z$) (FeO$_4$ and FeO$_6$ stretching vibrations along the $z$-axis) \\ \hline
P(3) & 302 & 344  & $B_{2u}$($y$) (FeO$_6$ and FeO$_6$ bending vibrations within the $xy$-plane) \\ \hline
P(4) & 311 & 346  & $B_{3u}$($x$) (FeO$_6$ and FeO$_6$ bending vibrations within the $xy$-plane) \\ \hline
P(5) & 336 & 361  & $B_{2u}$($y$) (FeO$_4$ and FeO$_6$ bending vibrations)                      \\ \hline
P(6) & 577 & 592  & $B_{2u}$($y$) (FeO$_4$ stretching vibration within the $xy$-plane)          \\ \hline
P(7) & 589 & 604  & $B_{3u}$($x$) (FeO$_4$ stretching vibration within the $xy$-plane)          \\ \hline
P(8) & {\bf 620}  & 619 & $B_{1u}$($z$) (FeO$_4$ and FeO$_6$ stretching vibrations along the $z$-axis)           \\ \hline
P(9) & {\bf 629}  & 621 & $B_{1u}$($z$) (FeO$_4$ and FeO$_6$ stretching vibrations along the $z$-axis) \\ \hline
\end{tabular}
}
\end{table}

\begin{figure}[t]
  \vspace*{-0.3 cm}%
  \centerline{\includegraphics[width=4.0 in]{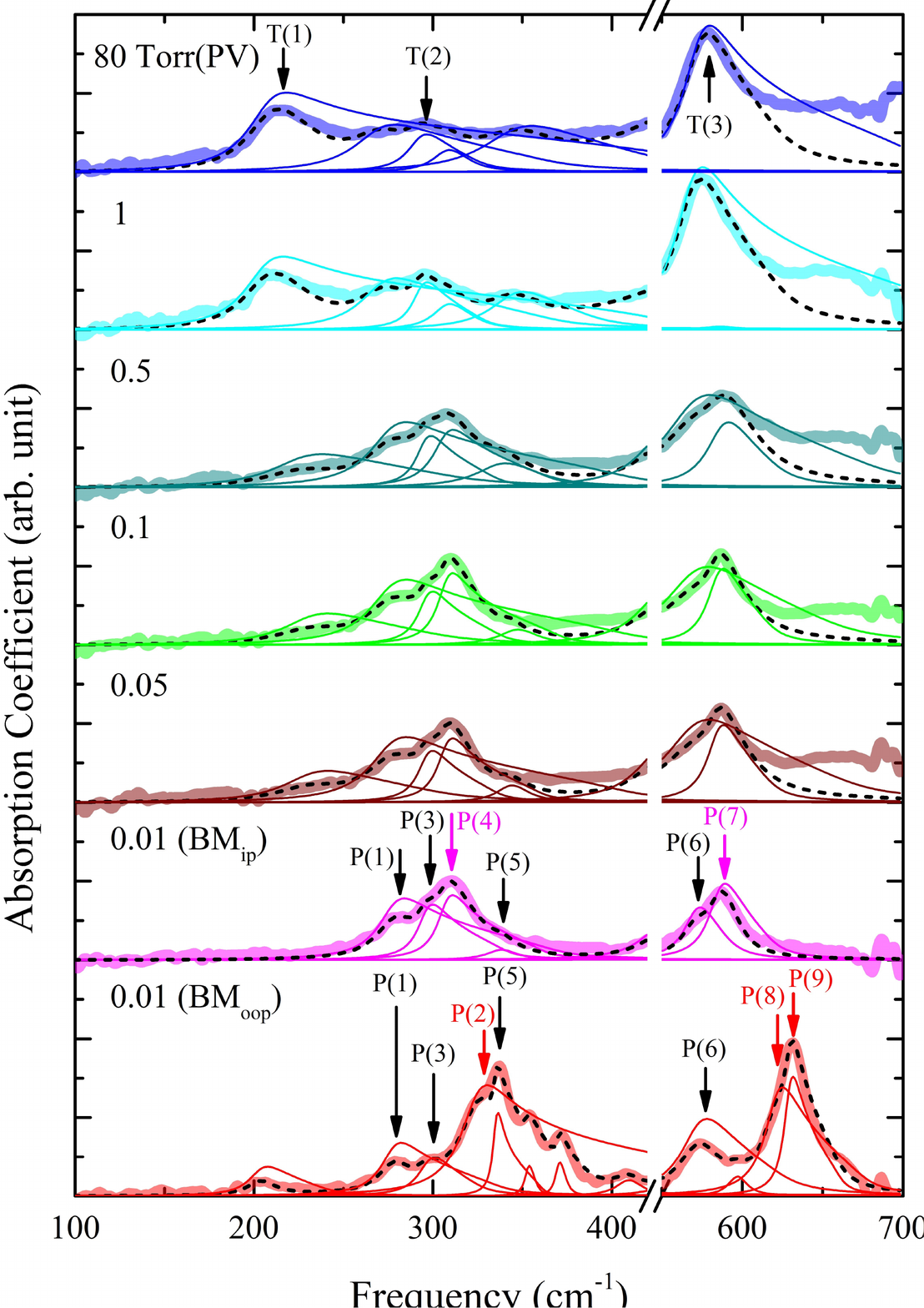}}%
  \vspace*{-0.3cm}%
\caption{Absorption coefficient spectra of all our SFO thin films at various oxygen vacancy levels. The thick lines represent measured data, and the thin lines represent separate phonon modes. The thick dashed black lines are the fits to the data. The phonons are assigned and labeled as P(1-9) for BM and T(1-3) for PV.}
 \label{fig6}
\end{figure}

Fig. 6(a) shows the absorption coefficient spectra of all the SFO films, which were obtained from the effective transmittance spectra. The BM$_{\mathrm{oop}}$ sample exhibits distinct phonon features compared with the other samples, because we post-annealed the BM$_{\mathrm{ip}}$ under various OPPs. As the OPP increased, the phonon modes show a systematic evolution until OPP = 0.5 Torr, then an abrupt structural change in between OPP = 0.5 and 1 Torr, and then another systematic evolution up to OPP = 80 Torr and above. There might be a structural crossover in between OPP = 0.5 and 1 Torr. We assigned selected nine phonon modes of BM structures, which are relatively strong and accessible with our experimental data, and labeled them as P(1-9) according to our shell-model calculations (refer to Fig. 3). The assigned modes are indicated with solid vertical lines in Fig. 3 and can be classified into two groups corresponding to Fe-O bending vibrations P(1, 3-5) and Fe-O stretching vibrations P(2, 6-9), which agree well with previous reports on other BM materials \cite{Tenailleau2005,Karlsson2008,Bielecki2014}. The three phonon modes of the PV phase are labeled as T(1-3).

The P(1) mode is observed in both the BM$_{\mathrm{oop}}$ and BM$_{\mathrm{ip}}$ absorption spectra. As the oxygen vacancy level decreases, this phonon mode softens, and its spectral weight is suppressed in the PV phases (OPP = 1 and 80 Torr). According to our calculation, the P(2) mode can be observed only in BM$_{\mathrm{oop}}$ since it oscillates only along the $z$-axis. We note that the assigned resonance frequency of the P(2) mode is relatively high compared with the assigned resonance frequencies of P(3)and P(4) (refer to Table III). The P(3) (344 cm$^{-1}$) can be seen in both BM$_{\mathrm{ip}}$ and BM$_{\mathrm{oop}}$ and has almost the same calculated resonance frequency as P(2) (343 cm$^{-1}$). P(3) exists even in the PV phase. The P(4) mode is observed only in the BM$_{\mathrm{ip}}$, and as the OPP increases, the spectral weight of this phonon decreases significantly. The P(5) mode can be observed in the both BM$_{\mathrm{oop}}$ and BM$_{\mathrm{ip}}$ samples. We note that the P(5) mode of the Bm$_{\mathrm{oop}}$ shows a stronger oscillation strength than that of BM$_{\mathrm{ip}}$ sample. According to the calculations the P(5) modes in both BM$_{\mathrm{oop}}$ and BM$_{\mathrm{ip}}$ vibrate along the $y$-axis and their spectral weights are similar to each other as shown in Fig. 3. There are several shoulder peaks near this mode on the higher-frequency side in the BM$_{\mathrm{oop}}$ sample. These shoulder peaks do not appear in the calculated phonon spectra and may come from the non-cubic crystal symmetry of the BM$_{\mathrm{oop}}$ sample \cite{Tenailleau2005,Karlsson2008,Bielecki2014}. P(6) is observed in both the BM$_{\mathrm{oop}}$ and BM$_{\mathrm{ip}}$ samples. As the oxygen vacancy level is reduced, this phonon mode becomes stronger, and when the BM sample is transformed into the PV structure, this phonon mode evolves to the T(3) mode. The P(7) phonon is only observed in the BM$_{\mathrm{ip}}$ sample, and as the OPP increases, it loses its spectral weight and exhibits a slight red-shift. When the structure transforms into the PV phase, the P(7) mode merges with the P(6) mode and evolves into the T(3) mode of the PV phase. The P(8) and P(9) modes are only observed in the BM$_{\mathrm{oop}}$ and are related to the vibration of apical oxygen in the octahedron.

In the PV phase samples (1, 80 Torr), we observed not only the T(1-3) phonon modes, but also three BM phonons (P(1,3, and 4)), suggesting that the PV phase sample contained a majority of the PV phase and a minority of the BM$_{\mathrm{ip}}$ phase. On the other hand, the 0.5 Torr sample shows the T(1) phonon, while its overall spectral shape is similar to BM$_{\mathrm{ip}}$, which can be explained by the fact that this sample is a mixed phase with a majority of the BM$_{\mathrm{ip}}$ phase. The mixed phase of the 0.5 Torr sample is consistent with previously reported XRD results\cite{Khare2017}. Remarkably, according to a previous study \cite{Khare2017}, the SFO$_{3-x}$ thin film exhibits intermediate and homogeneous phases during the topotactic phase transition from BM to PV at elevated temperatures owing to the itinerant oxygen. However, at room temperature, the film is stabilized as a mixed phase of BM and PV. Our phonon spectra show the same trends, providing more detailed structural information about our SFO thin films compared with the electronic structure analyses \cite{Dagotto2005,Mizusaki1992,Jeen2013t}. We note that P(2), P(8) and P(9) are observed only in the  BM$_{\mathrm{oop}}$ sample since the oscillations involved are along only the $z$-direction and their relative resonance frequencies are higher than those of other observed phonons compared with corresponding calculated resonance frequencies (refer to Table III). We speculate that these anomalous behaviors of P(2), P(8) and P(9) modes are related to the apical oxygen instability.

\begin{figure}[t]
  \vspace*{-1.0 cm}%
  \centerline{\includegraphics[width=7.0 in]{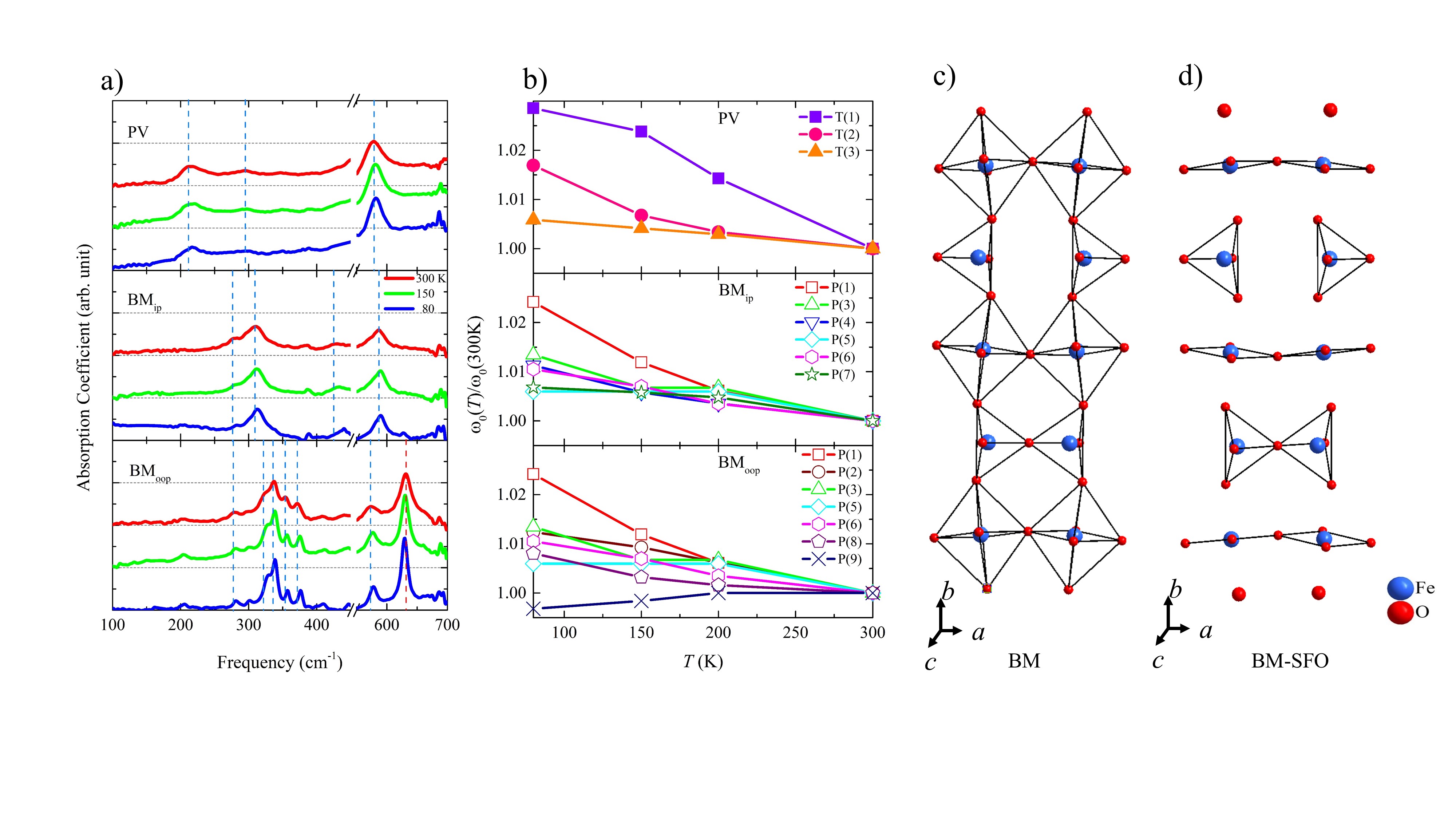}}%
  \vspace*{-2.0cm}%
\caption{(a) Temperature-dependent absorption coefficient spectra of three representative samples. The blue (blue-shift) and red (red-shift) vertical dashed lines are plotted to clearly illustrate the phonon shift. (b) Temperature-dependent $\omega_0$($T$)/$\omega_0$(300K) is displayed to illustrate the phonon shift. With cooling, the P(9) shows a red-shift, while others exhibit blue-shifts. (c) The normal BM crystal structure. The A atoms are not shown, for clarity. (d) BM crystal structure of SFO. The strontium atoms are not shown, for clarity. Here we drew (c) and (d) by referring to Fig. 7(a) of Ref. [14].}
 \label{fig7}
\end{figure}

Fig. 7(a) shows the absorption spectra of our three representative samples in the BM$_{\mathrm{ip}}$, BM$_{\mathrm{oop}}$ (OPP = 0.01 Torr), and PV (OPP = 80 Torr) phases at three selected temperatures. As the temperature decreases, the phonon modes become sharper, as the thermal smearing effects are reduced. At low temperatures, the lattice anharmonicity causes thermal contraction of the lattice, leading to a blue-shift of the phonon frequencies. We included the vertical dashed lines to illustrate the temperature-dependent frequency shifts. We observe that all the phonon modes of interest are blue-shifted (marked with the blue dashed lines), except for the phonon P(9) mode at $\sim$629 cm$^{-1}$ in the BM$_{\mathrm{oop}}$ sample, which is marked with a red dashed line. To depict the phonon mode shift more clearly, we display the $\omega_0$($T$)/$\omega_0$(300K) of all the assigned phonons in Fig. 7(b), where $\omega_0$($T$) represents the center frequency of each phonon at the temperature, $T$. Only the P(9) mode exhibits a red-shift by 0.3\%; the other phonons undergo a blue-shift by 0.6-3.0\%. This anomalous red-shift is related to the instability of the apical oxygen in the BM$_{\mathrm{ip}}$ structure \cite{Fleury1976,Lobo2007,Nakanishi1982}.

It has been reported that, in the BM structure of SrFeO$_{2.5}$, the tetrahedral ordering may distort the octahedra, resulting in the instability of the apical oxygen of the octahedra. This instability causes the apical oxygens to be displaced away from the central Fe ion in the FeO$_6$ octahedron and towards the FeO$_4$ tetrahedron \cite{Parsons2009,Paulus2008,Glamazda2015,Haruta2011,Piovano2015}. Because of these displacements, the typical alternating tetrahedral and octahedral layers are relaxed into a structure resembling that of SrFeO$_2$ with infinite (FeO$_2$)$_\infty$ layers and tetrahedral units between the layers as shown in Fig. 7(c) and 7(d). In this situation, the apical oxygens may experience two different bonding force fields along the symmetric axis of the octahedron: one is from the Fe ion in the FeO$_4$ tetrahedron unit, and the other from the Fe ion in the (FeO$_2$)$_\infty$ layers. We note that we did not clearly observe an additional phonon mode, which may be caused by the two different bonding force fields. The apical oxygen instability can cause the displacement of the apical oxygen from its original position in the octahedral FeO$_6$ and increases the bonding length between the apical oxygen and the Fe ion. Because of the greater bonding length, the bonding strength between the apical oxygen and the iron ion in the octahedron is significantly reduced. This bonding strength reduction may result in a stronger effect on the frequency of the related phonon (P(9)) than the unit cell volume contraction upon cooling, which may allow us to explain the observed abnormal red-shift. This is our speculation; it is not possible to visualize the detailed structural changes with the temperature based on optical spectra alone.

Inoue \textit{et al.} \cite{Inoue2010} reported that in BM, there are two oxygen diffusion pathways, which are highly anisotropic: diffusion within each tetrahedral or octahedral layer and diffusion through the tetrahedral and octahedral layer. Between theses, the diffusion within the layers is higher. This is supported by our experimental results; i.e., the apical oxygen in the octahedron is instable, and the bonding force with the Fe is smaller than the equatorial one. Thus, the apical oxygen in the octahedron is more likely to participate in the ionic conduction. Moreover, the instability inherent to the BM SFO appears to trigger the low temperature oxygen mobility, which has been suggested by Paulus \textit{et al.} \cite{Paulus2008}.

\section{Conclusion}

We prepared SrFeO$_{2.5}$ epitaxial thin film samples (BM$_{\mathrm{oop}}$ and BM$_{\mathrm{ip}}$) on an STO substrate using PLE. By treating the BM$_{\mathrm{oop}}$ samples with a post-annealing process at different OPPs, a series of SFO$_{3-x}$ thin films were obtained. From measured reflectance spectra in the FIR region, we extracted the absorption coefficient spectra of all our samples. In the absorption coefficient spectra, we observed characteristic phonon modes of each sample and compared them with the results of shell-model calculations. We found that a sample treated at OPP = 0.5 Torr was an inhomogeneous mixed phase of BM$_{\mathrm{ip}}$ and PV phases with BM$_{\mathrm{ip}}$ phase dominancy at room temperature, whereas samples treated at OPP = 1 and 80 Torr were a mixed phase with PV phase dominancy at room temperature. We also found that the P(9) phonon mode is related to the Fe-O stretching and undergoes an abnormal red-shift with cooling. We attribute this red-shift to the instability of the apical oxygen in the octahedron. It is not yet completely understood how the apical oxygen instability is related to the ionic conduction mechanism. However, our results provide the important clue that the instability of the apical oxygens can trigger the ionic conduction, which was proposed by Paulus \textit{et al.} \cite{Paulus2008}. This information enhances our understanding of the ionic conduction mechanism in these material systems. Furthermore, it provides better insight for designing the high ionic conducting materials.

%
%
\acknowledgments J.H. acknowledges financial support from the National Research Foundation of Korea (NRFK Grant No. 2017R1A2B4007387). This work was also supported by NRF-2017R1A2B4011083 (A.K. \& W.S.C.).

\bibliographystyle{apsrev4-1}
\bibliography{SFOPAPER}

\end{document}